\documentclass[aps,prl,reprint,twocolumn,floatfix,superscriptaddress,longbibliography]{revtex4-1}

\usepackage{lipsum}  

\usepackage[T1]{fontenc}
\usepackage{amsfonts, amsmath, amssymb}
\usepackage{graphicx}
\usepackage{dcolumn}
\usepackage{bm}
\usepackage[normalem]{ulem}
\usepackage{color}
\usepackage[utf8]{inputenc}
\usepackage[colorlinks,citecolor=blue,linkcolor=blue,urlcolor=blue]{hyperref}
\usepackage{xcolor}

\usepackage{braket}
\usepackage[caption=false]{subfig}
\graphicspath{{img/}}
\usepackage{color}
\usepackage{soul}
\usepackage{mathtools}
\usepackage{float}
\usepackage{bbold}

\newcommand{\vect}[1]{\boldsymbol{#1}}
\renewcommand{\vec}[1]{\vect{#1}}
\newcommand{\fhat}{\hat{f}}
\newcommand{\atrain}{\mathcal{A}_{\mathrm{train}}}
\newcommand{\atest}{\mathcal{A}_{\mathrm{test}}}

\def\rme{{\rm {e}}}
\def\rmi{{\rm {i}}}
\def\d{{\rm {d}}}

\newcommand{\be}{\begin{equation}}
\newcommand{\ee}{\end{equation}}

\newcommand{\norm}[1]{\left\lVert#1\right\rVert}

\newcommand{\hhat}{\hat{H}}

\newcommand{\id}{\mathbb{1}}

\newcommand{\ntrain}{{N_\mathrm{train}}}
\newcommand{\ntest}{{N_\mathrm{test}}}

\newcommand{\hc}{\mathrm{H.c.}}

\newcommand{\phihat}{\hat{\phi}}

\newcommand{\diag}{\mathrm{diag}}
\newcommand{\aaa}{\hat{a}}
\newcommand{\daaa}{\hat{a}^\dagger}
\newcommand{\mhat}{\hat{m}}
\newcommand{\xmu}{x^{\mu}}

\newcommand{\rhohat}{\hat{\rho}}

\newcommand{\bbb}{\hat{b}}
\newcommand{\dbbb}{\hat{b}^\dagger}

\begin{document}
\title{Machine learning via relativity-inspired quantum dynamics}

\author{Zejian Li}
\author{Valentin Heyraud}
\author{Kaelan Donatella}
\author{Zakari Denis}
\author{Cristiano Ciuti}
\affiliation{Universit\'{e} Paris Cit\'{e}, CNRS, Laboratoire Mat\'{e}riaux et Ph\'{e}nom\`{e}nes Quantiques (MPQ), 75013 Paris, France}

\begin{abstract}
We present a machine-learning scheme based on the relativistic dynamics of a quantum system, namely a quantum detector inside a cavity resonator. An equivalent analog model can be realized for example in a circuit QED platform subject to properly modulated driving fields. We consider a reservoir-computing scheme where the input data are embedded in the modulation of the system (equivalent to the acceleration of the relativistic object) and the output data are obtained by linear combinations of measured observables. As an illustrative example, we have simulated such a relativistic quantum machine for a challenging classification task, showing a very large enhancement of the accuracy in the relativistic regime. Using kernel-machine theory, we show that in the relativistic regime the task-independent expressivity is dramatically magnified with respect to the Newtonian regime.

\end{abstract}

\date{\today}

\maketitle

{\it Introduction.---}
Among several approaches for the conception of quantum computing devices, the field of relativistic quantum information~\cite{Peres_2004,Mann_2012} has also emerged. It has been demonstrated that non-inertial motion, or, via the equivalence principle, gravitational fields, can be used to generate quantum gates. Recent theoretical works have demonstrated that a non-uniformly accelerated cavity can generate two-mode squeezing ~\cite{bruschi_relativistic_2013} and cluster states~\cite{bruschi_towards_2016} for continuous-variable quantum computing~\cite{Braunstein_2005}. In a complementary scenario, a cavity remains inertial, but hosts accelerated detectors. Also for this configuration it has been shown that universal single-qubit rotations can be performed~\cite{martin-martinez_processing_2013}. While all the existing proposals for relativistic quantum computing require a very challenging control of mechanical motion, the corresponding models can be however synthesized in artificial platforms \cite{johansson2009dynamical,del_rey_simulating_2012,felicetti_relativistic_2015} such as those based on circuit QED \cite{CircuitQED_RMP} or trapped ions \cite{trapped_ions_RMP}.

In recent years, \textit{reservoir computing} has emerged as an appealing paradigm of information processing~\cite{Tanaka_2019}. This framework consists in approximating a target function by feeding its arguments as an input of a {\it reservoir}, whose dynamics nonlinearly maps the data into a high dimensional space. The resulting output data are then fed into a parametrized linear transformation to yield a trial function. These parameters are finally optimized through supervised-learning. This architecture makes the computational resources involved in the training process relatively modest. This has led to proposals and realizations in diverse platforms, including free-space
optics~\cite{sande2017,sunada2020,pierangeli2021}, photonics~\cite{vandoorne2014,denis-lecoarer2018}, nonlinear polariton lattices~\cite{opala2019,ballarini2020,mirek2021}, memristors~\cite{kulkarni2012,du2017} and beyond~\cite{boyn2017,nakane2018,markovic2019,marcucci2020}. Very recently, such an approach has been explored in a quantum context~\cite{markovic2020b,2021arXiv211110956A}, with applications in quantum metrology~\cite{ghosh2019a,ghosh2020}, quantum-state control~\cite{ghosh2019,ghosh2021,krisnanda2021} and image recognition~\cite{xu2021,2022arXiv220412192H}. 
Although it was long thought that a strong nonlinearity of the equations of motion was an essential element of reservoir computing, recent works have shown great performances relying on systems with almost no intrinsic nonlinearity, namely 
by exploiting the nonlinearity of the measurement~\cite{dong2020,rafayelyan2020,pierangeli2021} or drawing links with approximate kernel evaluation~\cite{saade2016,ohana2020,PhysRevApplied.17.034077}. 

In this letter, we present a reservoir-computing scheme exploiting the relativistic motion of a quantum system. We consider a paradigmatic model describing a quantum detector (atom) moving inside a cavity resonator. The relativistic dynamics can be simulated by an analog system such as a circuit QED platform with a tailored modulation of driving fields~\cite{del_rey_simulating_2012}. We explore the dynamics where the input data are embedded into the system by modulating the acceleration of the detector and measured output observables are then fed to a trainable linear classifier. By evaluating task-independent figures of merit, we demonstrate that the expressivity of our machine-learning protocol is dramatically enhanced in the relativistic regime. Moreover, we provide an illustrative example with a challenging classification task.
 
\begin{figure*}[t!]
    \centering
    \includegraphics[width=\linewidth]{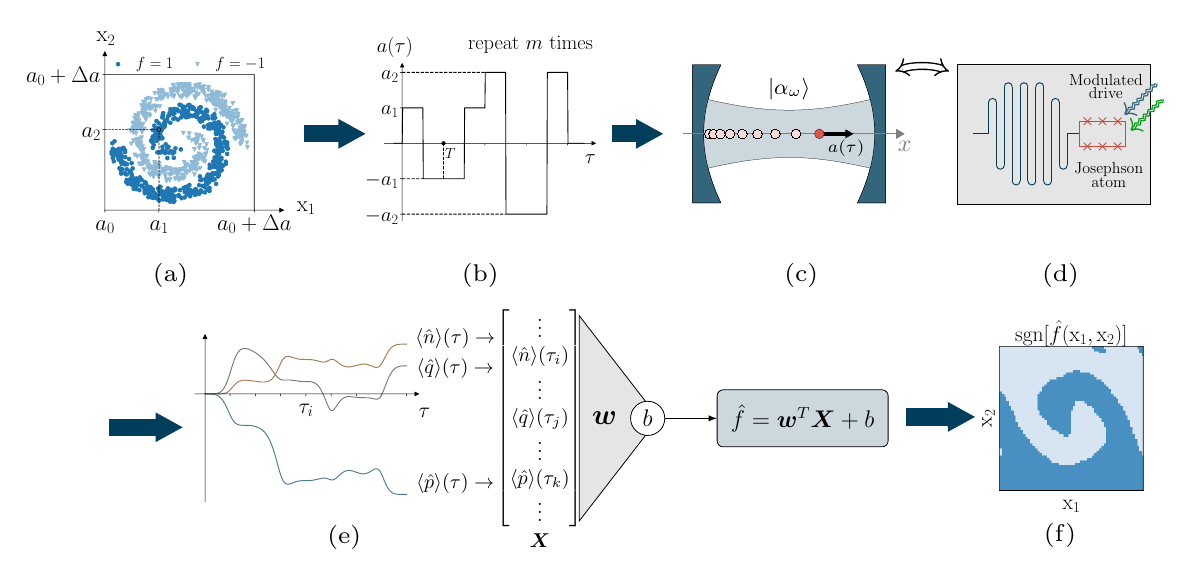}
    \caption{Scheme of the relativistic reservoir-computing protocol. (a) Each point $\vect{\mathrm{x}}=(\mathrm{x}_1,\mathrm{x}_2)$ of the dataset is linearly mapped to acceleration values $(a_1,a_2)$ according to Eq.~\eqref{eq:input_map}. (b) The acceleration values are used to construct a piece-wise constant acceleration profile $a(\tau)$. (c) The quantum detector, initially at rest in the cavity prepared in a single-mode coherent state, undergoes non-inertial motion with proper acceleration $a(\tau)$. (d) Analog circuit QED system where the analogous of the proper acceleration is controlled by modulated driving fields.  (e) Observables of the detector are measured at different times giving the feature vector $\vec{X}$ and the affine trial function $\hat{f}=\vec{w}^T\vec{X} + b$. (f) The classification result is predicted by $\mathrm{sgn}[\hat{f}(\mathrm{x}_1,\mathrm{x}_2)]$.}
    \label{fig:schema}
\end{figure*}

{\it Relativistic quantum model.---}
Let us consider the model describing a quantum harmonic detector with proper frequency $\Omega$, minimally coupled to a quantum field $\phihat$ inside an optical cavity. In the interaction picture, the corresponding Hamiltonian takes the Unruh-Dewitt form~\cite{brown_detectors_2013,Unruh_1984}
     $\hhat(\tau) = \lambda\mhat(\tau)\phihat[\xmu(\tau)]$,
where $\tau$ is the proper time of the detector, $\lambda$ is the coupling constant, and $\mhat(\tau) = \bbb\rme^{-\rmi\Omega \tau} + \dbbb \rme^{\rmi \Omega\tau}$ depends on the detector annihilation (creation) operator $\bbb$ ($\hat{b}^{\dagger}$). Finally,  $\xmu(\tau) = (t(\tau),x(\tau))$ is the world line of the detector in the $1{+}1$D Minkowski spacetime. We will adopt the metric $\eta_{\mu\nu}=\diag(+1,-1)$ and natural units such that $\hbar=c=1$. 
For a multi-mode cavity with perfectly reflecting mirrors~\cite{martin-martinez_processing_2013,ahmadzadegan_measuring_2014},
\begin{align}\label{eq:fullham}
    \hhat(\tau) &= \lambda \sum_{n=1}^{\infty}\dfrac{\sin[k_n x(\tau)]}{\sqrt{L\omega_n}}\times \\ &\left( \bbb\aaa_n\rme^{-\rmi[\Omega\tau+\omega_n t(\tau)]}+\bbb\daaa_n\rme^{-\rmi[\Omega\tau-\omega_n t(\tau)]} \right)
    + \hc,\nonumber
\end{align}
where $\omega_n=k_n=n\pi/L$ and $L$ is the cavity length.
The mode operators (denoting $\aaa_0\equiv\bbb$ for the detector) satisfy bosonic commutation relations  $[\aaa_n,\daaa_m] = \delta_{nm}$.
Both rotating and counter-rotating terms are present and contribute in the non-inertial regime~\cite{martin-martinez_processing_2013}.

Let us prepare the cavity in a single-mode coherent state whose frequency is resonant with that of the detector. Let us also consider the detector initially in its ground state $\rhohat_{0,(a)}=|0_a\rangle\langle 0_a|$.  The density matrix then reads:
\begin{equation}\begin{aligned}
    \rhohat_0 = \rhohat_{0,(a)}\otimes |\alpha_{\omega_i}\rangle\langle\alpha_{\omega_i}|\bigotimes_{j\neq i}|0_{\omega_j}\rangle\langle 0_{\omega_j}|.
\end{aligned}\end{equation}
For a given $\xmu(\tau)$, the time evolution of the density matrix is given by
\begin{equation}\begin{aligned}
    \dfrac{\d\rhohat(\tau)}{\d \tau} = -\rmi[\hhat(\tau),\rhohat(\tau)].
\end{aligned}\end{equation}
Since all the considered modes are bosonic and the Hamiltonian is quadratic, the Gaussianity of the initial state is preserved during the evolution. The dynamics of $\rhohat(\tau)$ can therefore be solved exactly using the covariance-matrix formalism for Gaussian states~\cite{brown_detectors_2013,PhysRevA.37.3028} [see Supplementary Material (SM) \cite{Note1}].

\footnotetext{See Supplemental Material at <link>.}

{\it Circuit QED analog implementation.---} 
As shown in the literature \cite{del_rey_simulating_2012}, the Hamiltonian \eqref{eq:fullham} can be implemented on a circuit QED platform consisting of an artificial Josephson atom \footnote{The anharmonicity of a Josephson atom can be made arbitrarily small by replacing a junction by a chain of junctions, as the anharmonicity scales as $1/N^2$ where $N$ is the number of junctions} coupled to a multi-mode transmission line resonator [see Fig.~\ref{fig:schema}(d)] \cite{del_rey_simulating_2012}. For a single mode, the Hamiltonian reads
\begin{equation}
    \hhat^{\mathrm{QED}}(\tau)=\omega_0\daaa\aaa + \epsilon\dbbb\bbb+ \eta\zeta(\tau)\dbbb\bbb\\
    + g(\dbbb+\bbb)(\daaa+\aaa),
\end{equation}
where $\zeta(\tau)$ is the sum of four driving fields containing two different tones and two non-adiabatically modulated phases (see SM \cite{Note1}). To get the equivalent of the acceleration range considered in this work, frequencies $\omega_0$ and $\epsilon$ can be taken in the standard GHz range while $g$, $\Omega$ and the phase modulation rates are in the MHz range (for more details, see SM \cite{Note1}). This driving scheme gives rise to the same interaction-picture Hamiltonian as Eq.~\eqref{eq:fullham} for a single mode, and can be similarly extended to multi-mode circuit-QED systems \cite{Kuzmin2019,PuertasMartnez2019,Mehta_2022}. While here we will report the results with a harmonic detector, we have also simulated the other extreme case where the Josephson atom is a {\it two-level system} and found similar results (see SM \cite{Note1}), showing that the detailed atom spectrum is not crucial.

{\it Reservoir-computing protocol.---}
Now, we can show how to exploit the relativistic system as a reservoir-computing model. The goal is to learn a nonlinear function $f(\vect{\mathrm{x}})$ of the input $\vect{\mathrm{x}}\equiv (\mathrm{x}_1,\mathrm{x}_2,...,\mathrm{x}_N)$ that follows some probability distribution  $\mu(\d\vect{\mathrm{x}})$ (see Fig. \ref{fig:schema}). Assuming that for every $i$ we have $\mathrm{x}_{i,\mathrm{min}}\leq \mathrm{x}_i \leq \mathrm{x}_{i,\mathrm{max}}$, we map them linearly to acceleration values in a fixed range between $a_0$ and $a_0+\Delta a$, namely
\begin{equation}\begin{aligned}\label{eq:input_map}
 \mathrm{x}_i \mapsto a_i = a_0 + \Delta a\times\dfrac{\mathrm{x}_i - \mathrm{x}_{i,\min}}{\mathrm{x}_{i,\max}-\mathrm{x}_{i,\min}}.
\end{aligned}\end{equation}
We then impose a piecewise constant proper acceleration $a(\tau)$ to the harmonic detector. The pieces have proper acceleration values $(a_1,-a_1,-a_1, a_1,\cdots,a_N,-a_N,-a_N,a_N)$ and each piece has a duration of $T/2$ in the proper frame of the detector, and we repeat this encoding sequence $m$ times. Assuming the detector to be initially at rest at $\xmu(\tau=0)=(t=0,x=0)$, this acceleration profile guarantees that at each instant $\tau=nT$, $n\in\mathbb{N}$, the detector is at rest, and that at $\tau=2nT$ it comes back to its original spatial position at $x=0$. Note that for a circuit QED implementation the modulation of the driving fields can directly control the analog of the proper acceleration with respect to the proper time $\tau$ (see SM \cite{Note1}).   The detector world line for a general  proper acceleration $a(\tau)$ is (see SM \cite{Note1} for a derivation)
\begin{equation}\begin{aligned}\label{eq:wl}
    x(\tau) = \int_0^\tau \d\tau' \sinh[\xi(\tau')]\,,~
    t(\tau) =\int_0^\tau \d\tau'\cosh[\xi(\tau')],
\end{aligned}\end{equation}
where $\xi(\tau) = \int_0^\tau\d\tau' a(\tau')$ is the rapidity~\cite{rindler1991introduction}. Instead, in the Newtonian case, the (unphysical) world line is simply
\begin{equation}\begin{aligned}\label{eq:newt}
    x_\mathrm{Newt}(\tau) = \int_0^\tau\d\tau'\xi(\tau')\,,~
    t_\mathrm{Newt}(\tau) = \tau.
\end{aligned}\end{equation}

Each input data point $\vect{\mathrm{x}}$ determines a single time evolution of the system $\rhohat(\tau)$.  We can then measure the detector at times $\tau_n = n\times \Delta T $ to obtain the expectation values~\footnote{We assume ensemble measurements for the quantum expectation values.} of the quadrature  operators $\hat{q} = (\bbb+\dbbb)/\sqrt{2}$, $\hat{p} = \rmi(\dbbb-\bbb)\sqrt{2}$ and of the number operator $\hat{n} = \dbbb\bbb$. The measurements are then collected into a feature vector $\vec{X}$ [see Fig.~\ref{fig:schema} (e)]. Finally, our trial function reads
\begin{equation}\begin{aligned}\label{eq:11}
\hat{f}(\vect{\mathrm{x}}) =\vec{w}^T\vec{X}(\vect{\mathrm{x}}) + b,
\end{aligned}\end{equation} where the weight $\vec{w}$ and bias $b$ are parameters to be optimized in order for $\hat{f}$ to approximate the target function $f$. To simplify the notation, in the following we will absorb $b$ into the vector $\vec{w}$ by appending a constant component $1$ to the  vector $\vec{X}$.
Due to the linear dependence of the trial function on the feature vector, its optimization can be done analytically.

{\it Results and discussion.--- }
As an illustrative example, we consider a non-trivial task: the two-spiral classification problem~\cite{lang1989learning}. The goal is to distinguish two interlocking spiral planar patterns. This task serves as a well-known benchmark for binary pattern classification that is considered hard for multi-layer perceptron models due to its complicated decision boundary \cite{Chien_Cheng_Yu}. The input data are the two coordinates of each point in the two-spiral pattern $\vect{\mathrm{x}}=(\mathrm{x}_1,\mathrm{x}_2)$. The task function $f$ to be learned is such that $f(\vect{\mathrm{x}}) = 1$ if the point belongs to the first spiral branch, and $f(\vect{\mathrm{x}}) = -1$ for the other branch [see Fig. \ref{fig:schema} (a)].
To train the model, we draw a train dataset of $\ntrain=4000$ sample points $\{\vect{\mathrm{x}}_{(1)},\vect{\mathrm{x}}_{(2)},\cdots,\vect{\mathrm{x}}_{(\ntrain)}\}$ with labels $y_i=f[\vect{\mathrm{x}}_{(i)}]$ and minimize the regularized least-square loss function. Although most classification problems are commonly treated with other losses \cite{hastie2013}, the training can be analytically performed under this choice of loss over the training set:
\begin{equation}\begin{aligned}
\mathcal{L}\left(\vec{w}\right) =\frac{1}{2\ntrain}\sum_{i=1}^{\ntrain}[y_i - \vec{w}^T\vec{X}(\vect{\mathrm{x}}_{(i)})]^2
+\frac{l}{2}\norm{\vec{w}}_2^2,
\label{eq:L2loss_with_reg}
\end{aligned}\end{equation}
where the last term is a regularization term to prevent overfitting.
Denoting $\vec{\Phi}$ the matrix whose $j$-th column is  $\vec{X}(\vec{\mathrm{x}}_{(j)})$, and $\vec{y}$ the column vector of the training labels $y_i$, the optimal weights are given by~
$
\vec{w}^{\star} = (\vec{\Phi}\vec{\Phi}^T + l \ntrain \id)^{-1}\vec{\Phi}\vec{y}.
\label{eq:optimal_weights}
$
The performance of the model is then evaluated on a test sample with $\ntest=1000$ points. We evaluate the classification accuracy $\atest$ on the test set as the fraction of correctly classified samples among $\ntest$. The training accuracy $\atrain$, which indicates how well the reservoir-computing model fits the training set, is defined analogously.
\begin{figure*}[t!]
    \centering
    \includegraphics[width=0.8\linewidth]{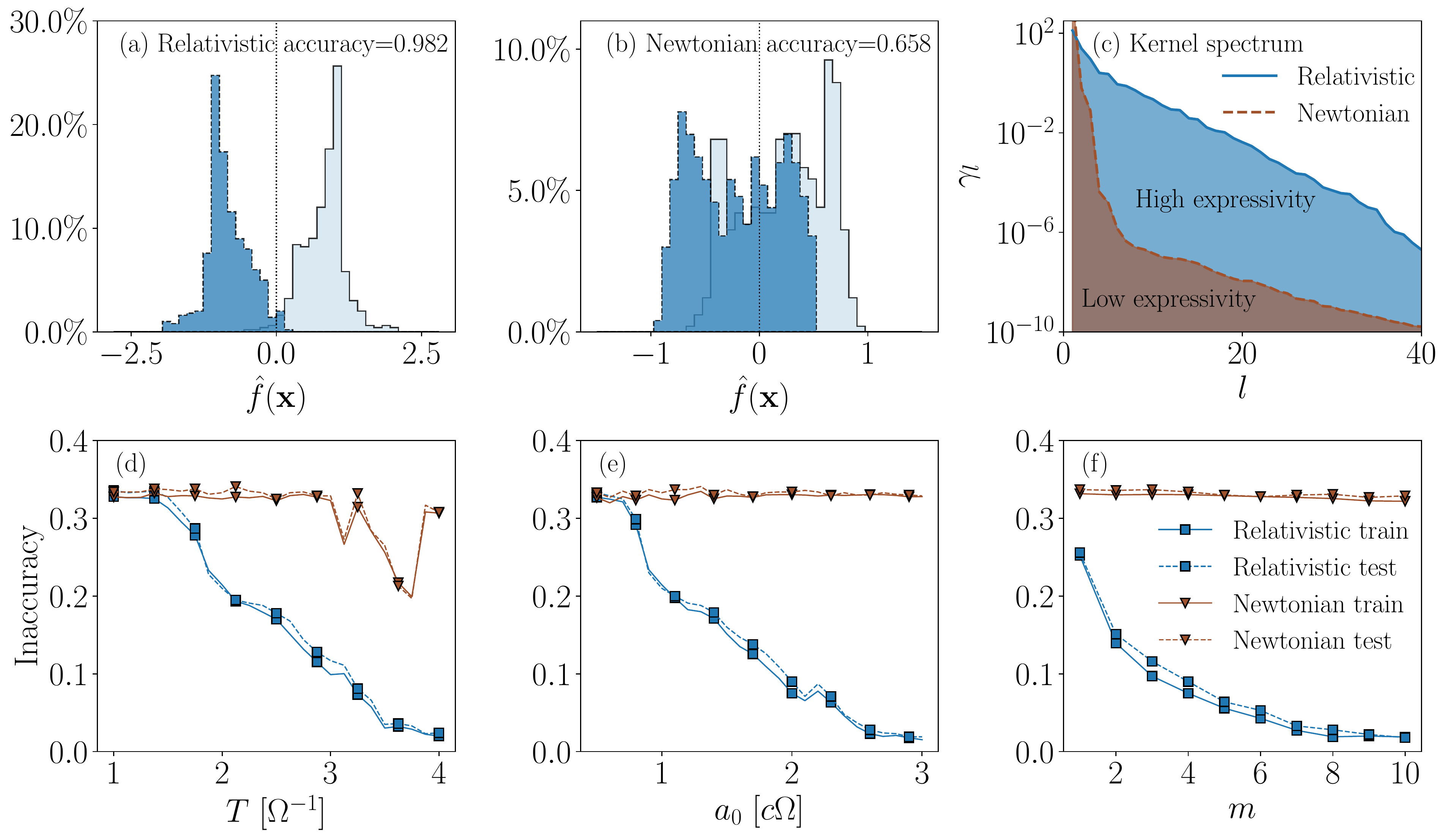}
    \caption{Figures of merit of the relativistic reservoir-computing protocol. (a) Light and dark histograms correspond to testing samples belonging to different spirals of the dataset. Parameters: $a_0=3$, $T=2$ and $m=4$.  (b) Same quantity plotted for the Newtonian model with same parameters. (c) The empirical kernel spectrum computed for the relativistic (solid line) and Newtonian (dashed line) models with same parameters. The first $40$ non-zero eigenvalues $\gamma_l$ are plotted in descending order. (d) Inaccuracy of the relativistic (triangles) and Newtonian (squares) models evaluated on both the training (solid lines) and testing (dashed lines) set, as a function of the acceleration time $T$. Parameters: $a_0=1$ and $m=4$. (e) Same quantities plotted as a function of the base acceleration $a_0$, for $T=2$ and $m=4$. (f) Same quantities plotted as a function of the number of repetitions $m$, for $a_0=2$ and $T=2$. Quantities are expressed in natural units, where the scale is fixed by the proper frequency of the atom $\Omega$.}
    \label{fig:results}
\end{figure*}
The transformation of the input $\vect{\mathrm{x}}\mapsto\vec{X}(\vect{\mathrm{x}})$ can be regarded as an embedding of the input from input space into some higher-dimensional feature space. This is best understood by introducing the  kernel function \cite{hastie2013}
 $   k(\vect{\mathrm{x}},\vect{\mathrm{x}}') = [\vec{X}(\vect{\mathrm{x}'})]^T\vec{X}(\vect{\mathrm{x}})$.
Under quite general assumptions, this kernel  can be diagonalized into an orthonormal~\footnote{with respect to the inner product on $L^2_\mu$: {$\langle f,h\rangle \equiv \int f(\vect{\mathrm{x}})h(\vect{\mathrm{x}})\mu(\d \vect{\mathrm{x}})$}, where $\mu(\d\vect{\mathrm{x}})$ denotes the probability distribution of the samples} set of eigenfunctions $\{\psi_i\}_i$ with positive eigenvalues $\{\gamma_i\}_i$~\cite{paulsen2016}:
\begin{equation}\begin{aligned}
    k(\vect{\mathrm{x}},\vect{\mathrm{x}}') &= \sum_i \gamma_i \psi_i(\vect{\mathrm{x}})\psi_i(\vect{\mathrm{x}}').
\end{aligned}\end{equation}

The set $\{\psi_i\}_i$ can be completed to be a basis of $L^2_\mu$~\footnote{The space of square-integrable functions with respect to the probability measure of the input distribution $\mu(\d\vect{\mathrm{x}})$.} with eigenfunctions associated with $\gamma_i=0$. Note that the kernel spectrum can be empirically computed~\cite{Williams01usingthe} by calculating the eigenvalues of the matrix $\vec{\Phi}\vec{\Phi}^T/\ntrain$. The trial function of Eq.~\eqref{eq:11} may be rewritten in the above kernel eigenbasis as
$\label{eq:17}
    \hat{f}(\vect{\mathrm{x}}) = \vect{\beta}^T\vect{\psi}(\vect{\mathrm{x}})$
where $\vec{\beta}$ is the weight vector to be optimized. The spectrum of the kernel contains crucial information as eigenfunctions with vanishingly small associated eigenvalues do not contribute to the expressivity of the reservoir~\cite{PhysRevApplied.17.034077}.  In what follows, we will use the kernel spectrum to assess the expressivity of the model. 

Throughout our simulations, we fixed the coupling constant to $\lambda=0.1$, the interval of measurement to $\Delta T = T/2$ and $\Delta a/a_0=0.1$. The detector's proper frequency is set to be resonant with the third cavity mode~\footnote{This is to ensure that the cavity is long enough for the atom to remain inside. For the circuit QED implementation this is not an important detail.} $\Omega=\omega_3$, the latter being initially in a coherent state $\lvert\alpha\rangle$ with $\alpha = 10\rmi$. We express all quantities in natural units with the scale fixed by $\Omega$. The regularization is set to $l=10^{-6}$; this is equivalent to having a measurement noise of variance $l$ in the observables~\cite{goodfellow2016}. Fig.~\ref{fig:results} (a) shows the distribution of testing samples in feature space, represented by $\hat{f}(\mathrm{x})$.
The same quantities are plotted in Fig.~\ref{fig:results} (b) in a non-relativistic setting, that is, considering Newtonian world lines [Eq. \eqref{eq:newt}]. As appears from Fig.~\ref{fig:results}(a) and (b), the relativistic model correctly separates the two classes with high accuracy.  By contrast, the system undergoing Newtonian dynamics exhibits a poor performance. The empirical kernel spectra of the two models are plotted in Fig.~\ref{fig:results} (c), where we show the first $40$ non-zero eigenvalues in descending order. The flatter distribution of the relativistic kernel spectrum implies that for a fixed cutoff threshold on the eigenvalues (or a fixed regularization~\cite{PhysRevApplied.17.034077}), it has more eigenfunctions with nonzero eigenvalues that can contribute to the expressivity of the trial function $\fhat$ in comparison with the Newtonian model. Importantly, this relativistically enhanced kernel expressivity associated to the dynamics is {\it task-independent} and explains the much higher accuracy achieved by the relativistic model for the specific two-spiral classification task.

In Fig.~\ref{fig:results} (d), we examine the impact of the acceleration time $T$ on the performance of the model. As $T$ increases, the inaccuracy ($1-\mathcal{A}$) of the relativistic model decreases to around $0$, whereas the performance of the Newtonian model remains poor.  This is consistent with the results of Fig.~\ref{fig:results} (e), where we vary the base acceleration $a_0$ for fixed $T$. Therein, we also found the inaccuracy of the relativistic model to be decreasing as a function of $a_0$ as the motion enters the relativistic regime, and a poor performance of the Newtonian model, which remains insensitive to $a_0$.

In Fig.~\ref{fig:results} (f) we study the effect of the number of repetitions $m$ of the encoding sequence on the performance. As we are taking measurements at a constant interval $\Delta T$, a larger value of $m$ allows for more features to be collected in the feature vector $\vec{X}$, improving the efficiency. By contrast, in the Newtonian setting, the supplementary features are close-to-linearly related to the previous ones, thus yielding a negligible improvement. The induced nonlinearity of the feature map associated to the dynamics of the relativistic reservoir ensures that the generated features remain non-trivial after many repetitions. The advantage of the relativistic model can be understood from  Eq.~\eqref{eq:fullham}. Indeed, as discussed in Ref.~\cite{martin-martinez_processing_2013}, the phases $\rme^{-\rmi[\Omega\tau\pm\omega_n t(\tau)]}$ depend non-trivially on $\tau$ due to the relativistic (time-dilation) effects, which  yields an input-dependent modulation of the cavity-detector resonance condition, absent in the Newtonian model, where one always has $t_{\mathrm{Newt}}(\tau)=\tau$.

{\it Conclusions.---}
We have shown how relativistic quantum dynamics can provide a dramatic enhancement of the expressive power for reservoir computing. Given that analogs of the considered relativistic quantum model can be implemented in state-of-the-art quantum platforms, such as superconducting circuits and trapped ions, our theoretical findings pave the way to relativity-inspired machine-learning protocols with enhanced capabilities. 

\begin{acknowledgements}
We acknowledge support by the FET FLAGSHIP Project PhoQuS (grant agreement ID: 820392) and by the French Projects NOMOS (ANR-18-CE24-0026),  and TRIANGLE (ANR-20-CE47-0011). We also thant the Ile de France region via the program SIRTEQ. This work was granted access to the HPC resources of TGCC under the allocation A0100512462 attributed by GENCI (Grand Equipement National de Calcul Intensif).
\end{acknowledgements}

\bibliography{bib}
\clearpage

\end{document}


\title{{\bf Supplementary Material for the article:}\\ ``Machine learning via relativity-inspired quantum dynamics''}

\author{Zejian Li}
\author{Valentin Heyraud}
\author{Kaelan Donatella}
\author{Zakari Denis}
\author{Cristiano Ciuti}
\affiliation{Universit\'{e} Paris Cit\'{e}, CNRS, Laboratoire Mat\'{e}riaux et Ph\'{e}nom\`{e}nes Quantiques (MPQ), 75013 Paris, France}

\date{\today}
\maketitle
In this document, we briefly summarize the Gaussian formalism adopted for calculating the time-evolution of the relativistic quantum system (Section~\ref{sec:gaussian}), the derivation of the detector world line (Section~\ref{app:wl}), the analog implementation of the relativistic model with circuit QED (Section~\ref{app:cqed}) and the simulation results with a qubit instead of a harmonic detector (Section~\ref{sec:qubit}).

\section{Gaussian formalism for the model with a harmonic detector}\label{sec:gaussian}
We denote the vector of bosonic mode operators by
\bea
\hat{\vec{\Psi}}=(\aaa_0,\aaa_1,\aaa_2,\cdots,\aaa_N,\daaa_0,\daaa_1,\daaa_2,\cdots,\daaa_N)^T,
\eea
that satisfies the commutation relation
\bea
    [\hat{\Psi}_i,\hat{\Psi}_j]=\Obold_{ij},
\eea
where 
\bea 
    \Obold = 
    \begin{bmatrix}
\mathbf{0} & \id\\
-\id & \mathbf{0}
\end{bmatrix} = -\Obold^T
\eea
is the symplectic form.
If the Hamiltonian can be written in the form of 
\bea
    \hhat = \hat{\vec{\Psi}}^T\mathbf{F}(t)\hat{\vec{\Psi}},
\eea
it then preserves the Gaussianity of states~\cite{PhysRevA.37.3028}.
The Heisenberg equations of motion can be written as
\bea 
    \dfrac{\d}{\d t}\hat{\vec{\Psi}} = -\rmi \Obold\Fsym(t)\hat{\vec{\Psi}},
\eea
where $\Fsym=\Fbold+\Fbold^T$.

If we define the propagator $\Sbold(t)$ via the relation
\bea
    \hat{\vec{\Psi}}(t) = \Sbold(t)\hat{\vec{\Psi}}(0),
\eea
it then satisfies the first order linear differential equation
\bea
        \dfrac{\d}{\d t}\Sbold(t) = -\rmi \Obold\Fsym(t)\Sbold(t)
\eea
with the initial condition $\Sbold(0)=\id$.
The evolution of the covariance matrix
\bea
    \sigma_{ij} = \langle\hat{\Psi}_i\hat{\Psi}_j\rangle - \langle\hat{\Psi}_i\rangle\langle\hat{\Psi}_j\rangle
\eea
is given by
\bea
    \sigbold(t) = \Sbold(t)\sigbold(0)\Sbold^T.
\eea
$\sigbold$ together with $\langle\hat{\vec{\Psi}}\rangle$ will completely specify a Gaussian state.

\section{Derivation of the world line}\label{app:wl}

We derive here the world line for an observer with time-dependent proper acceleration $a(\tau)$ in 1+1D Minkowski spacetime with metric $\eta_{\mu\nu}=\diag(+1,-1)$.

We parametrize the world line by the proper time $\xmu(\tau) = (t(\tau),x(\tau))$ and denote 
\bea
    u^\mu(\tau) =& \dfrac{\d}{\d \tau}\xmu(\tau) = (u^t(\tau),u^x(\tau)),\\
    a^\mu(\tau) =& \dfrac{\d}{\d\tau}u^\mu(\tau) = (a^t(\tau),a^x(\tau)).
\eea

From the definition of these quantities, we get
\bea
    u^\mu u_\mu =& 1 = (u^t)^2 - (u^x)^2,\\
    a^\mu u_\mu =& 0 = a^t u^t - a^x u^x,\\
    a^\mu a_\mu =& -a(\tau)^2 = (a^t)^2 - (a^x)^2.
\eea
It follows that
\bea
    a^x = \dfrac{\d u^x}{\d \tau} = a(\tau)\sqrt{1+(u^x)^2}.
\eea
Integrating from $\tau'=0$ to $\tau'=\tau$ gives
\bea
    u^x(\tau) =& \sinh[\xi(\tau)],\\
    \xi(\tau) =& \sinh^{-1}[u^x(\tau=0)] + \int_0^{\tau}\d\tau'a(\tau').
\eea
Integrating again gives the position. For the time, a similar treatment applies. We finally obtain:
\bea\label{eq:wlapp}
    x(\tau) =& x_0 + \int_0^\tau \d \tau' \sinh[\xi(\tau')],\\
    t(\tau) =& t_0 + \int_0^\tau\d \tau'\cosh[\xi(\tau')],
\eea
which reduces to the world-line equations in the main text for an observer initially at rest.
Note that for a constant acceleration, this gives the well-known Rindler observer's world line.

\section{implementation with circuit QED}\label{app:cqed}
We present a potential implementation of the proposed relativistic model on circuit QED platforms inspired by \cite{del_rey_simulating_2012}, which consists of a Josephson artificial atom with bosonic mode operator $\bbb$ (simulating the harmonic oscillator described in the main text) coupled to a microwave cavity in the strong-coupling regime. Denoting the microwave cavity mode operator by $\aaa$, the noninteracting Hamiltonian of the system is
\bea
    \hhat_0(\tau) = \omega_0\daaa\aaa+\epsilon\dbbb\bbb+\eta \zeta(\tau)\dbbb\bbb\, ,
\eea
where $\omega_0$ is the cavity bare frequency, $\epsilon$ is the energy of the artificial atom, and we assumed that the Josephson junction has negligible nonlinearity. This can be achieved for example by replacing a single Josephson junction with a sufficiently long chain of junctions. The opposite extreme case, where the Josephson atom is a two-level system (qubit), yields similar results, as revealed by corresponding simulations reported in Section~\ref{sec:qubit}. $\zeta(\tau)$ is a driving function that takes the following form~\footnote{Note that our driving term is different from that in~\cite{del_rey_simulating_2012}, as they considered simulating a quantum field in free space, while in the present work, the simulated quantum field is confined within a cavity with Dirichlet boundary conditions, resulting in different mode functions. }:
\bea
    \zeta(\tau)&=\dfrac{\d}{\d\tau} F(\tau)\, ,\\
    F(\tau)&=F_+(\tau)+F_-(\tau)\, ,
\eea
where
\bea\label{eq:drive}
    F_{\pm}&=\cos[\omega_{\pm}\tau\mp\theta_{\mp}(\tau)]-\cos[\omega_{\pm}\tau\mp\theta_{\pm}(\tau)]\, .
\eea
Assuming that the phases $\theta_\pm(\tau)$ are modulated slowly compared to the driving frequencies $\omega_\pm$, as will indeed be the case in what follows, the driving function $\zeta(\tau)$ can be well approximated by
\bea
    \zeta(\tau) \simeq& -\omega_+\sin[\omega_+\tau-\theta_-(\tau)] \\
    &+ \omega_+\sin[\omega_+\tau -\theta_+(\tau)] \\
    &- \omega_- \sin[\omega_-\tau+\theta_+(\tau)]\\
    &+ \omega_-\sin[\omega_-\tau+\theta_-(\tau)].
\eea
The interaction Hamiltonian in the Schrödinger picture is $\hhat_I=g(\dbbb+\bbb)(\aaa+\daaa)$. Passing to the interaction picture with respect to $\hhat_0(\tau)$ and assuming $\eta\ll 1$ in the driving term, we get
\bea
    \hhat_I(\tau) =& g[\dbbb\rme^{\rmi \epsilon\tau}\Gcal(\tau)+\hc](\aaa\rme^{-\rmi\omega_0\tau}+\hc)\,,\\
    \Gcal(\tau)=&\rme^{\rmi\eta F(\tau)}\simeq 1 + \rmi \eta F(\tau)\, .
\eea 
To simulate a harmonic oscillator with proper frequency $\Omega$ and world line $\xmu(\tau) = (t(\tau),x(\tau))$ coupled to the $n$th mode of a massless scalar field of frequency $\omega_n=k_n$ as considered in the main text, we now choose $\omega_{\pm} = \epsilon \pm \omega_0 -\Omega$ as the driving frequencies and $\theta_{\pm}(\tau) = \omega_n t(\tau)\pm k_n x(\tau)$ as the phase modulations. In the regime where  $\epsilon$, $\omega_0$, $|\epsilon\pm\omega_0|\gg g$, the interaction Hamiltonian becomes (keeping only slowly rotating terms)
\bea\label{eq:cqedHeff}
    \hhat_I(\tau)\simeq& g\eta \sin[k_n x(\tau)]\times\\
    &\bbb(\aaa\rme^{-\rmi[\Omega\tau+\omega_n t(\tau)]}+\daaa\rme^{-\rmi[\Omega\tau-\omega_n t(\tau)]})+\hc\, ,
\eea
which takes the form of the interaction Hamiltonian in the main text for a single mode of the quantum field. Note that since we always consider a single-mode coherent state as the field initial state in the main text, the main contribution to the dynamics of the harmonic oscillator comes uniquely from this mode, as one can verify using perturbation theory~\cite{martin-martinez_processing_2013}. We also checked numerically that a single-mode approximation for the quantum field is enough for obtaining accurate results for the simulations presented in the main text. Nonetheless, it is possible to simulate the full many-mode Hamiltonian by using multiple modes in the circuit QED microwave cavity. 

As considered in \cite{del_rey_simulating_2012}, the energy scales $\epsilon$ and $\omega_0$ for circuit QED are in the GHz regime, while $g$, $\Omega$ and $\omega_n$ can be on much slower time scales, such as in the MHz regime. 
The modulation rate of the phases $\dot{\theta}_\pm(\tau)$ can be expressed in terms of the simulated time-dependent acceleration $a(\tau)$ as [using the world line in Eq.~\eqref{eq:wlapp}]
\bea\label{eq:phaseder}
    \dot{\theta}_\pm (\tau) &= \dfrac{\d}{\d\tau}[\omega_n t(\tau)\pm k_n x(\tau)]\\
    &= \omega_n \cosh[\xi(\tau)] \pm k_n \sinh[\xi(\tau)]\\
    &= \omega_n \cosh\left[\int_0^\tau\d \tau' a(\tau')\right] \pm k_n \sinh\left[\int_0^\tau\d \tau' a(\tau')\right]\, .
\eea

Let us consider a typical world line studied in the main text, for example with $a_0=2$, $\Delta a/a_0=0.1$, $T=2$ and $\omega_n=\Omega$ [the values used in Fig. 2(f) of the main text] in the units fixed by $\Omega$. Then, we have $\dot{\theta}_{\pm}(\tau)\lesssim 10\Omega$, meaning that the phases in the driving \eqref{eq:drive} need to be modulated at roughly the same timescale as $\Omega$, in the MHz band, which is much slower than the circuit QED timescales and should be experimentally feasible. 

\begin{figure}[htbp]
    \centering
    \includegraphics[width=0.9\linewidth]{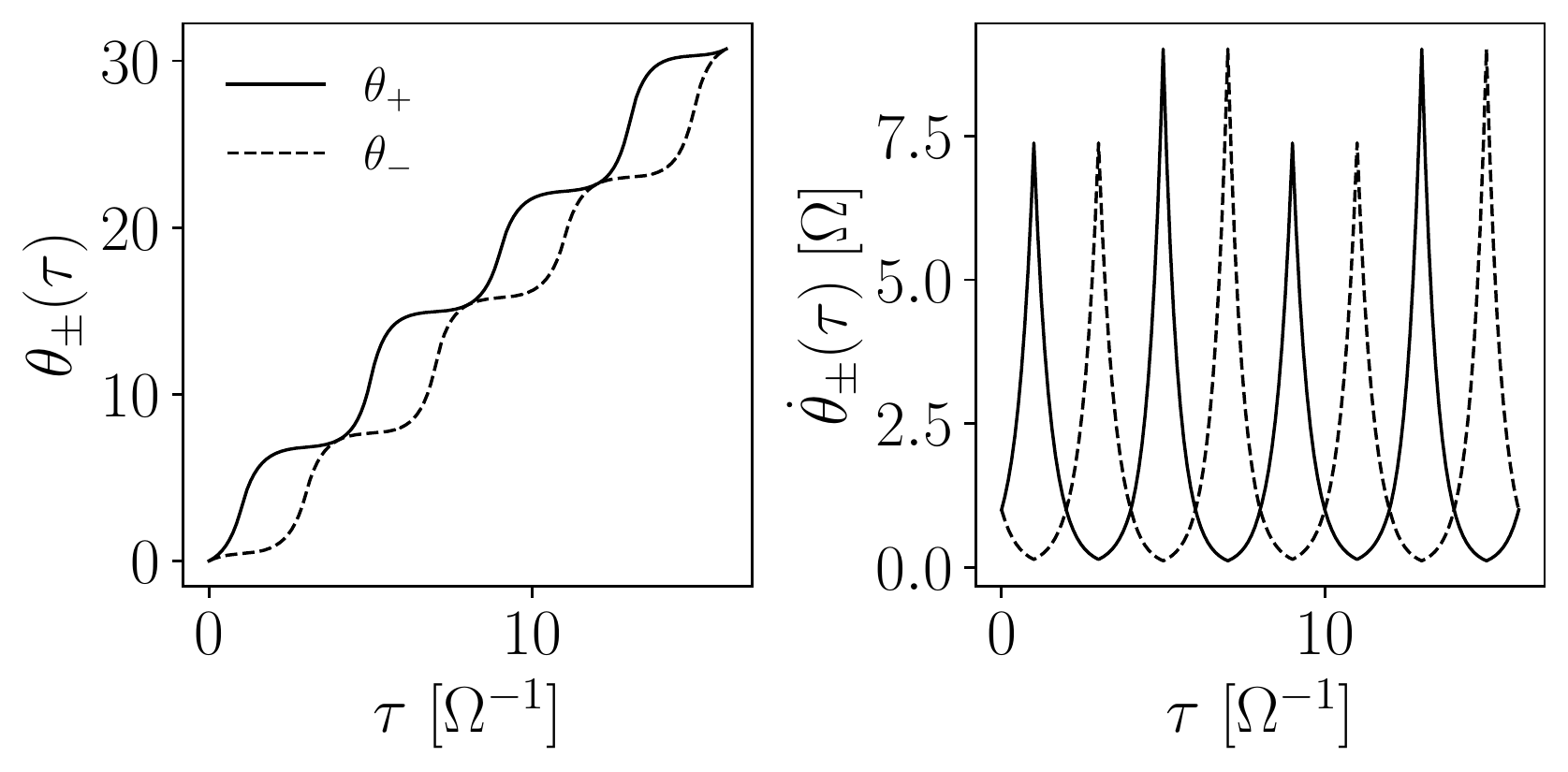}
    \caption{The phase modulations $\theta_\pm(\tau)$ (left panel) and the corresponding rates $\dot\theta_\pm(\tau)$ that provide the desired simulation of the accelerated motion, where $\Omega=\SI{1}{\MHz}$. Note that modulation rates are in the regime of $\dot\theta_\pm(\tau)\lesssim 10\Omega=\SI{10}{\MHz}$.}
    \label{fig:phases}
\end{figure}

Finally, let us consider a concrete example of typical parameter values of the analog circuit QED system. Let the parameters of the system be $\omega_0=\SI{1}{\GHz}$, $\epsilon=\SI{1.1}{\GHz}$, $\Omega=\SI{1}{MHz}$, $g=10/\sqrt{3\pi}\,\si{\MHz}\simeq \SI{3.3}{\MHz}$, $\eta=0.01$. The driving frequencies are then $\omega_+=\SI{2.099}{\GHz}$ and $\omega_-=\SI{0.099}{\GHz}$. This simulates the harmonic detector coupled to the $n=3$ mode of the quantum field (with $\Omega=\omega_n=k_n$ and $\lambda=0.1$) as considered in the main text. To simulate the acceleration sequence described in the main text in the case of $a_0=2$, $\Delta a/a_0=0.1$ and $T=2$, the required phase modulations $\theta_\pm(\tau)$ as well as their rates $\dot\theta(\tau)$, given by Eq.~\eqref{eq:phaseder}, are plotted in Fig.~\ref{fig:phases}.

\section{Results with a qubit instead of a harmonic oscillator}\label{sec:qubit}

\begin{figure}[htbp]
    \centering
    \includegraphics[width=0.9\linewidth]{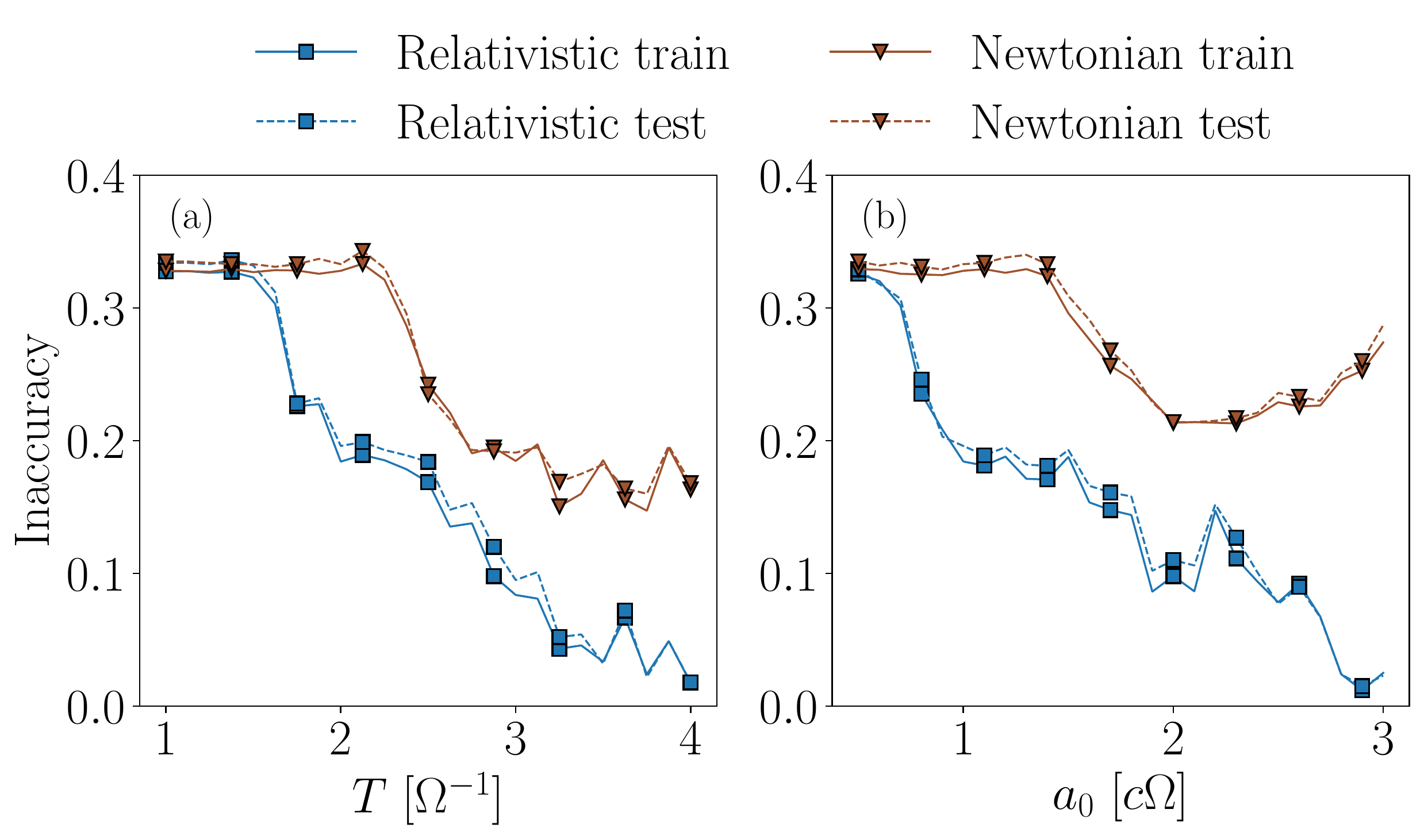}
    \caption{Performances of the reservoir-computing model considered in the main text, but with a qubit replacing the harmonic oscillator. Same parameters as Fig. 2(d) and (e) in the main text respectively, showing very similar results.}
    \label{fig:qubit}
\end{figure}

We report here the simulation results when we replace the harmonic detector with a qubit (two-level atom initially in its ground state) for the same parameters considered in the main text. To model the configuration with the qubit, we have to replace the bosonic mode operator $\bbb$ with the Pauli operator $\sigmam$ in the Hamiltonian. Since the Gaussian formalism can no longer be applied, we assumed a single-mode approximation for the quantum field (considering only the mode that is initially in the coherent state and in resonance with the proper frequency of the qubit), which matches the exact form of the single-mode circuit-QED Hamiltonian in Eq.~\eqref{eq:cqedHeff}. The feature vector now contains the expectation values of the operators that are respectively analogous to the bosonic occupation number and the quadratures, namely $\sigmap\sigmam$, $(\sigmam+\sigmap)/\sqrt{2}$ and $\rmi(\sigmap-\sigmam)\sqrt{2}$.  This is equivalent to measuring the Pauli operators $\sigmaz$, $\sigmax$ and $\sigmay$ respectively. The equivalent of Fig. 2(d) and (e) in the main text are presented in Fig.~\ref{fig:qubit}(a) and (b) for the qubit model. We recover results similar to the case of the harmonic detector. Note that the Newtonian model has a slightly improved, yet still very poor performance, which can be ascribed to the additional nonlinearity provided by the qubit. These results clearly show that the details of the spectrum of the detector are not crucial for the expressive power of the relativistic quantum dynamics.
 
\newpage

\bibliography{bib}